\documentclass[checkin,showpacs,aps,prx]{revtex4}

\newcommand{\ba}{\begin{eqnarray}}
\newcommand{\ea}{\end{eqnarray}}

\usepackage{graphicx,color}

\newcommand{\ce}{$\rm C_{60}$}
\newcommand{\cm}{$\rm C_{60}$ }

\newcommand{\be}{\begin{equation}}
\newcommand{\ee}{\end{equation}}
\newcommand{\la}{\langle}

\newcommand{\ra}{\rangle}
\newcommand{\et}{{\it et al. }}
\newcommand{\ete}{{\it et al.}}

\def\prl{{ Phys. Rev. Lett. }}

\def\prb{{ Phys. Rev. B }}
\def\pra{{ Phys. Rev. A }}

\begin{document}

\title{Imaging superatomic molecular orbitals in a ${\bf
    C_{60}}$ molecule through four 800-nm photons}

 \author{G. P. Zhang,$^{1,*}$ H. P. Zhu,$^{1,2}$ Y. H. Bai$^3$, J.
   Bonacum$^1$,
   X. S. Wu$^{2}$, and Thomas F. George$^4$}

 \affiliation{$^1$Department of Physics, Indiana State University, Terre
   Haute, Indiana 47809, USA}

 \affiliation{$^2$Laboratory of Solid State Microstructures and School
   of Physics, Nanjing University, Nanjing 210093, China}

 \affiliation{$^3$Office of Information Technology, Indiana State
   University, Terre Haute, Indiana 47809, USA}

\affiliation{$^4$
   Office of the Chancellor and Center for Nanoscience \\Departments
   of Chemistry \& Biochemistry and Physics \& Astronomy \\University
   of Missouri-St. Louis, St. Louis, MO 63121, USA } \date{\today}

\begin{abstract}
{Superatomic molecular orbitals (SAMO) in \cm are ideal building
  blocks for functional nanostructures.  However, imaging them
  spatially in the gas phase has been unsuccessful. It is found
  experimentally that if \cm is excited by an 800-nm laser, the
  photoelectron casts an anisotropic velocity image, but the image
  becomes isotropic if excited at a 400-nm wavelength. This diffuse
  image difference has been attributed to electron thermal ionization,
  but more recent experiments (800 nm) reveal a clear non-diffuse
  image superimposed on the diffuse image, whose origin remains a
  mystery.  Here we show that the non-diffuse anisotropic image is the
  precursor of the $f$ SAMO.  We predict that four 800-nm photons can
  directly access the $1f$ SAMO, and with one more photon, can image
  the orbital, with the photoelectron angular distribution having two
  maxima at 0$^\circ$ and 180$^\circ$ and two humps
  separated by 56.5$^\circ$.  Since two 400-nm photons only resonantly
  excite the spherical $1s$ SAMO and four 800-nm photon excite the
  anisotropic $1f$ SAMO, our finding gives a natural explanation of the
  non-diffuse image difference, complementing the thermal scenario.  }
\end{abstract}


\maketitle

\section{Introduction}

In nanoscience, no other cluster has ever garnered more attention than
\cm \cite{kroto}. Such a highly symmetric molecule allows a high
electron delocalization, enabling ultrafast dynamics
\cite{ultrafast,prb07}, strong nonlinear optical responses
\cite{nlo,prl12}, and high harmonic generation
\cite{prl05,john}. Recently, employing scanning tunneling microscopy
(STM), Feng \et \cite{feng} discovered a group of unusually large
orbitals both inside and outside of the \cm cage, which are both
beautiful and surprising. These orbitals, called superatomic molecular
orbitals (SAMOs), have attracted immediate attention \cite{steve,
  johansson12}.  Spatially, they bear a close resemblance to their
atomic counterparts, but with a much larger radius. This motivated Roy
and coworkers \cite{roy} to design nanoscale atoms for solid-state
chemistry.  Figure \ref{fig1} shows a $1f$ orbital, where one sees a
distinctive anisotropy characteristic of $f$ orbitals.  However,
imaging such high-lying orbitals becomes increasingly difficult for
STM, because of the strong overlap between the orbitals of \cm and
those of the substrates.

The velocity-map imaging technique \cite{eppink,bordas} does not have
this problem, since it works in the gas phase. Figure \ref{fig1}
schematically shows that after the laser strikes \ce, the
photoelectrons with different velocities, after several stages of
accelerating electrode plates, cast an image on the phosphor screen,
which is captured by the CCD camera. This image carries the orbital
information. Doing so, quite surprisingly, neither Johansson \et
\cite{johansson12} nor Kjellberg \et \cite{kjellberg1} detected any
image similar to superorbitals. Instead, they reported a seemingly
irrelevant observation: The electron image is anisotropic if excited
by an 800-nm laser, but is isotropic with a 400-nm laser. How could it
be possible that a similar observation in ion yields (ellipticity
dependence) \cite{hertel09,shc09} occurs for the electrons as well?
Huismans \et \cite{huismans} further showed that the anisotropy
increases with the laser intensity, and they assigned it to the
accumulation of highly structured but slightly different angular
distributions of the ground states, while others assign the diffuse
part of the photoelectron spectrum to thermal electron ionization
\cite{kjellberg1} or a ``kick'' from the laser field
\cite{johansson12}, with an explanation presented in the second paper
of Ref. \cite{johansson12}.  Huismans \et emphasized that a fully
satisfactory theory requires a complete {\it ab initio}
calculation. Such a calculation is now available \cite{johansson}, but
these studies did not address the difference between the 800-nm and
400-nm excitations. Very recently, Li and coworkers experimentally
showed a distinct and non-diffuse image superimposed on the diffuse
anisotropic background \cite{li}. These non-diffuse features can not
easily be explained by the thermal electron ionization, since the
Boltzmann-like distribution is unlikely to yield an image with
intensities only concentrated at six spatial locations.  We wonder how
such  nondiffuse images are formed, besides the origin in the thermal
electron ionization.

In this paper, we demonstrate theoretically that SAMOs in \cm in
general, and the $1f$ orbital in particular, are accessible to the
velocity map imaging technique.  Our first important prediction is
that the energy gap between the $1f$ orbital and HOMO matches the
four-photon energy of an 800-nm laser pulse. By successively absorbing
four photons, electrons in the HOMO can be excited into the $1f$
orbital; by absorbing one extra photon, they are cleared of the
ionization potential energy of 7.58 eV, and can then be detected by the
velocity map imaging. We further predict that the angular resolved
spectrum has a distinctive $1f$ orbital feature, with two local
maxima, separated by 56.5$^\circ$. There are two reasons why the $1f$
orbital can be effectively probed. First, the multiphoton excitation
suppresses an otherwise strong contribution from nonSAMOs. Second,
although energetically $d$ and $g$ SAMOS are in the vicinity of the
$1f$ orbital, due to the selection rule, transitions to those orbitals
are not possible. Such a detection scheme is quite generic.  By
slightly tuning the photon energy of the laser beam, it is also
possible to observe $2p$, $3p$ and $1h$ orbitals. If \cm is excited by
two 400-nm photons, the strongest transition is between HOMO-1 and
nonSAMOs, but this transition is off-resonant by 0.62 eV. On the other
hand, although the transition matrix element product is small for the
transition from HOMO-2 to 1\textit{s} SAMO, this transition is nearly resonant,
and it is likely that the 1\textit{s} SAMO is excited, consistent with
Ref. \cite{johansson12}.  Therefore, our study suggests a new and
alternative explanation to the anisotropic and non-diffuse image
\cite{li}, and demonstrates the new power of multiphoton emission for
investigating SAMOs in fullerenes.

\section{Theoretical formalism}

We employ two complementary first-principles methods within the
density functional theory: the basis function-based methods
(Gaussian09 \cite{gaussian} and VASP \cite{vasp}) and the real grid
mesh method (Octopus) \cite{octopus}. To build a case to directly
probe SAMOs in \ce, we first
investigate the density functionals' effect on the energy gap between
the HOMO and $1f$ SAMO for a
group of six different functionals, within the same basis 6-311+G(d).
Figure \ref{fig2}(a) shows that the functional has a substantial
impact on the energy gap $\Delta E$ (see the filled red
circles). Since the ionization potential (IP) is at 7.58 eV (the
dashed green line) and $1f$ SAMO must be a few eV below IP, some popular
functionals such as B3LYP and CAM-B3LYP predict the results too large
even in the single-particle limit. This result is rigorous since it
does not involve the complicated photoexcitation and ionization. This
convinces us that the SWVN/LDA functional is reliable, at least within
the single-particle limit.

Next, we employ the same SWVN/LDA functional but increase the basis
function size by adding more diffuse functions \cite{steve} to the
Gaussian basis. When we use the aug-cc-pvtz basis, we find that
$\Delta E$ drops to 7.136 eV (see the empty red circle in
Fig. \ref{fig2}(a)).  We then put a layer of 60 fictitious hydrogen
atoms at a radius of 9 $\rm \AA$ from the center of \ce. These atoms
have no charge and only serve as the centers for additional basis
functions.  The carbon atoms have the 6-311+G(d) basis, while those
fictitious hydrogen atoms have aug-cc-pV6Z, which contains $7s$,
$6p$, $5d$, $4f$, $3g$ and $2h$ Gaussian primitive functions. Doing
so, we find that the LDA gap $\Delta E_{1f,LDA}$ decreases from 7.506
eV (without fictitious atoms) to 6.957 eV (with fictitious
atoms), with the net change as large as 0.549 eV (compare the first
two filled circles of Fig. \ref{fig2}(a)).  It is convincing that the
basis functions strongly affect the accuracy of the energy levels of
SAMOs. However, if we continue to use the Gaussian primitives, the
calculation becomes increasingly demanding. We must pursue a different
approach.

To gradually eliminate the basis function effect, we employ a grid
mesh-based method as implemented in Octopus \cite{octopus}. An
important advantage is that the real grid is a balanced approach and
treats SAMOs and nonSAMOs on an equal footing.  Figure \ref{fig2}(a)
shows that as $r$ increases (the bottom horizontal axis represents
$r$), $\Delta E$ gradually converges to 6.337 eV (see the squares at
the bottom of the figure). We then fit the data to $\Delta E=\Delta
E_0+\Delta E_1/r$ and find $\Delta E_0=6.015$ eV and $\Delta
E_1=10.0707$ eV/$\rm \AA$. $\Delta E_0=6.015$ eV is considered as a
lower limit for this gap. As an independent check, we employ a
planewave basis function as implemented in the VASP code \cite{vasp},
where we place \cm in a big fcc supercell with two respective lattice
constants of $a$ = 21 $\rm \AA$ and 25 $\rm \AA$, and two planewave cutoffs of
500 and 600 eV. The results are shown in Fig. \ref{fig2}(a) (see the
plus signs). The VASP result is 6.36 eV, which is considered as an
upper limit. If we average these two limits, we expect that this gap
settles down at 6.2 eV, which precisely matches the four-photon energy
of a laser pulse of wavelength 800 nm.  Figure \ref{fig2}(b) shows the
energy spectrum of SAMOs referenced to the HOMO. For an 800-nm laser,
five photons are needed to reach the IP (see the thick dashed green
line on the top). The fourth photon has access to the SAMOs which are
congested in a narrow window. Figure \ref{fig2}(c) zooms in on the
SAMOs. Consistent with Feng's VASP result \cite{feng}, the gap between
the $s$ SAMO and LUMO is 3.25 eV, but our interest is in higher SAMOs.
This energy gap becomes an outstanding case to judge the quality of
the two most popular density functionals, LDA versus B3LYP.  If the
LDA result is correct, then an 800-nm laser can detect $1f$ SAMO
experimentally.  This is the first testable case for experiments.

\section{Two-photon versus four-photon excitation}
Regardless of the experimental finding, such a multiphoton excitation
requires a stronger laser.  Would it be possible to do the same job
with two UV photons as Johansson \et \cite{johansson} did?  In order
to have a successive optical transition from the occupied states to
unoccupied states, the accumulative product of matrix elements of each
transition must be different from zero.  Different from prior studies,
we compute the transition matrix elements between the HOMOs, SAMOs and
nonSAMOs by integrating \be \la i|{\bf D}|j \ra
=\int^{\infty}_{-\infty} d\tau \psi_i({\bf r}) {\bf r} \psi_f({\bf
  r}), \ee where $\psi_i({\bf r})$ is the Kohn-Sham wavefunction, and
${\bf r}$ is the electron coordinate. For the frequency-doubled case
\cite{johansson}, at least two photons are needed to reach the $1f$
SAMO. Thus, we compute the product of two transition matrix elements,
$\la occ|{\bf D}|i\ra\la i|{\bf D}|j\ra$, for all the relevant
transitions around this two-photon energy. The results are
surprising. Although there are lots of states energetically
accessible, only a very few actually contribute. Figure
\ref{4photon}(a) shows that for two-400-nm-photon photoexcitation, two
groups of nonSAMOs have the largest product of the transition matrix
elements, followed by $f$, $s$, $d$, $g$ and $h$ SAMOs. Although the
nonSAMOs dominate the product of the matrix elements, their energies
are off-resonant from two-400 nm photoexcitation. Specifically, the
energy for two 400-nm photons is 6.2 eV, but the excitation from
HOMO-1 to the nonSAMOs is 6.83 eV, off by 0.63 eV, which greatly
reduces the excitation probability into the nonSAMOs.  But for the
1\textit{s} SAMO, although its matrix product is small, the excitation energy
from HOMO-2 to 1\textit{s} SAMO is 6.17 eV, nearly resonant with the two-400 nm
photon energy of 6.2 eV. Therefore, during 400-nm excitation, the 1\textit{s} SAMO is excited strongly. This finding is also consistent with
Johansson's assignment of their first peak to $1s$ SAMO
\cite{johansson}. Since the 1\textit{s} SAMO is spherical, this explains why at
400-nm excitation the photoelectron image is isotropic, besides the
contribution due to the diffuse mechanism as discussed above.

In the same spirit, we investigate the four-photon process.
Four-photon excitations build upon four successive transitions, with
the product of four transition moments as $ \la i|{\bf D}|j\ra\la
j|{\bf D}|k\ra \la k|{\bf D}|l\ra\la l|{\bf D}|m\ra $, where $|i\ra$
is the HOMO, $|m\ra$ are the SAMOs, and the others are intermediate
states.  Figure \ref{4photon}(b) plots the products for all the states
with large contributions.  Our results are insightful. First, the
transition matrix element strictly obeys the dipole selection rule and
has the correct parity. We see that only SAMOs with odd angular
momentum quantum numbers have nonzero products. For instance, $s$ and
$d$ SAMOs have no transition. Second, the SAMO dominates the
four-photon transitions. The largest product is from the $1h$ SAMO,
and the second largest is from the $1f$ SAMO, followed by the $2p$
and $3p$ SAMOs.  It is likely that they compete with the $1f$ SAMO,
but which one dominates depends sensitively on the actual photon
energy, a finding that allows one to probe various SAMOs by tuning the
laser wavelength.  We also notice that the nonSAMO contribution is
still sizable. Therefore, to observe SAMOs, future experiments must
steer the laser wavelength away from those nonSAMOs energetically and
use an optimal laser intensity (a laser too strong would populate
other nonSAMOs). For an 800-nm laser, $1f$ is strongly excited, both
in terms of the transition matrix elements and the transition energy.
Since its shape is highly anisotropic (see the far right schematic in
Fig. \ref{fsamo}), this explains why experimentally Li and coworkers \cite{li}
find that the photoelectron image is non-diffusive and anisotropic, and
more importantly has fine structures superimposed on the diffuse image
seen before \cite{johansson12,kjellberg1}.  To test our results
quantitatively, we compute the kinetic energy of the electron.  To
eject this electron from $1f$ SAMO, one more photon is needed to clear
the ionization potential energy IP = 7.58 eV. This leaves the ionized
electron with a kinetic energy of $5 h\nu-\textrm{IP}=0.2$ eV.  Sure
enough, Fig. 2(e) of Ref. \cite{johansson} indeed shows a peak at
0.2 eV, but this important finding did not catch the attention of the
investigators, as their emphasis was on the 400-nm results.

\section{Discussion}

\begin{figure}
\includegraphics[angle=0,width=14cm]{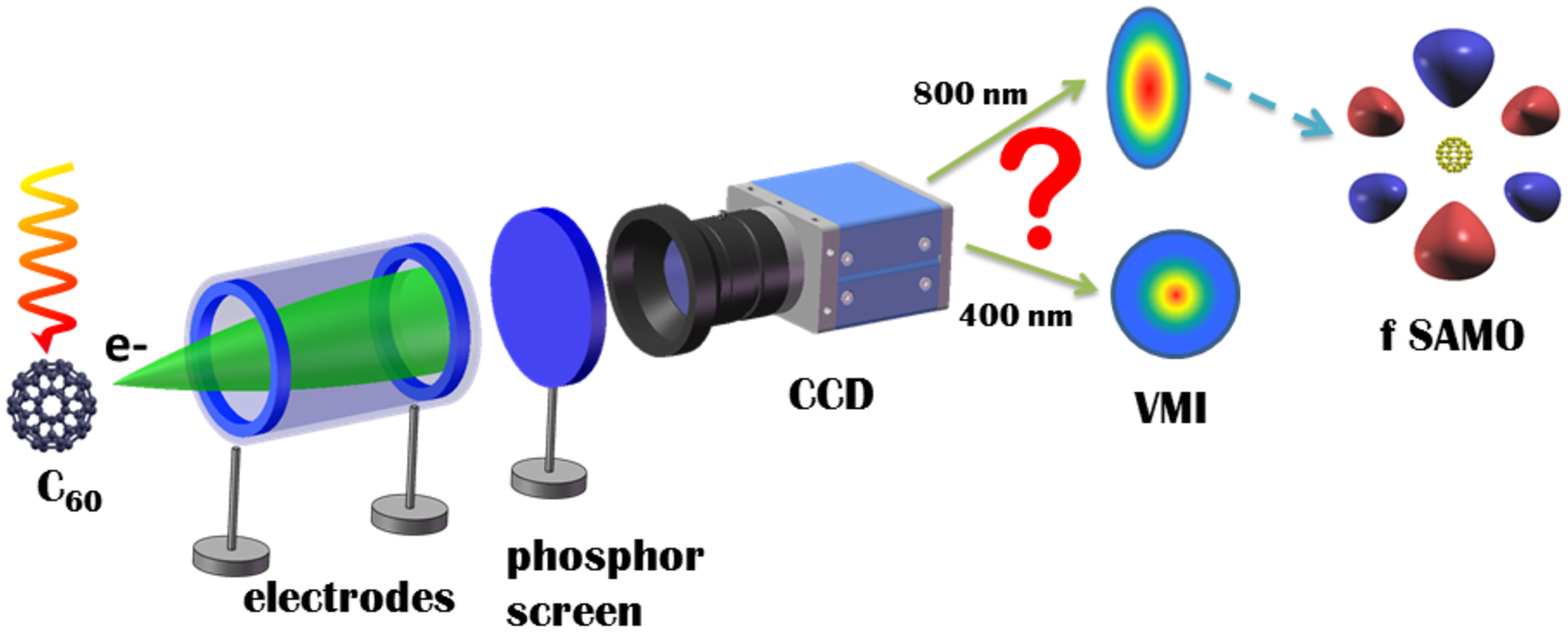}
\caption{Proposed experimental setup of velocity map
  imaging. Photoelectrons from \cm pass through electrode plates and
  form images on a phosphor screen, which are in turn captured by a
  CCD camera.  Depending on whether an 800-nm or 400-nm laser is used,
  the image is anisotropic or isotropic. The diffuse origin of this
  image difference is attributed to the thermal photoelectron
  emission, but this can not explain the non-diffuse features seen in
  a recent experiment \cite{li}.  Far right: For convenience of
  viewing, we rotate the $f$ superatomic molecular orbital (SAMO) by
  90$^\circ$. }
\label{fig1}
\label{fsamo}
\end{figure}

\begin{figure}
\includegraphics[angle=270,width=14cm]{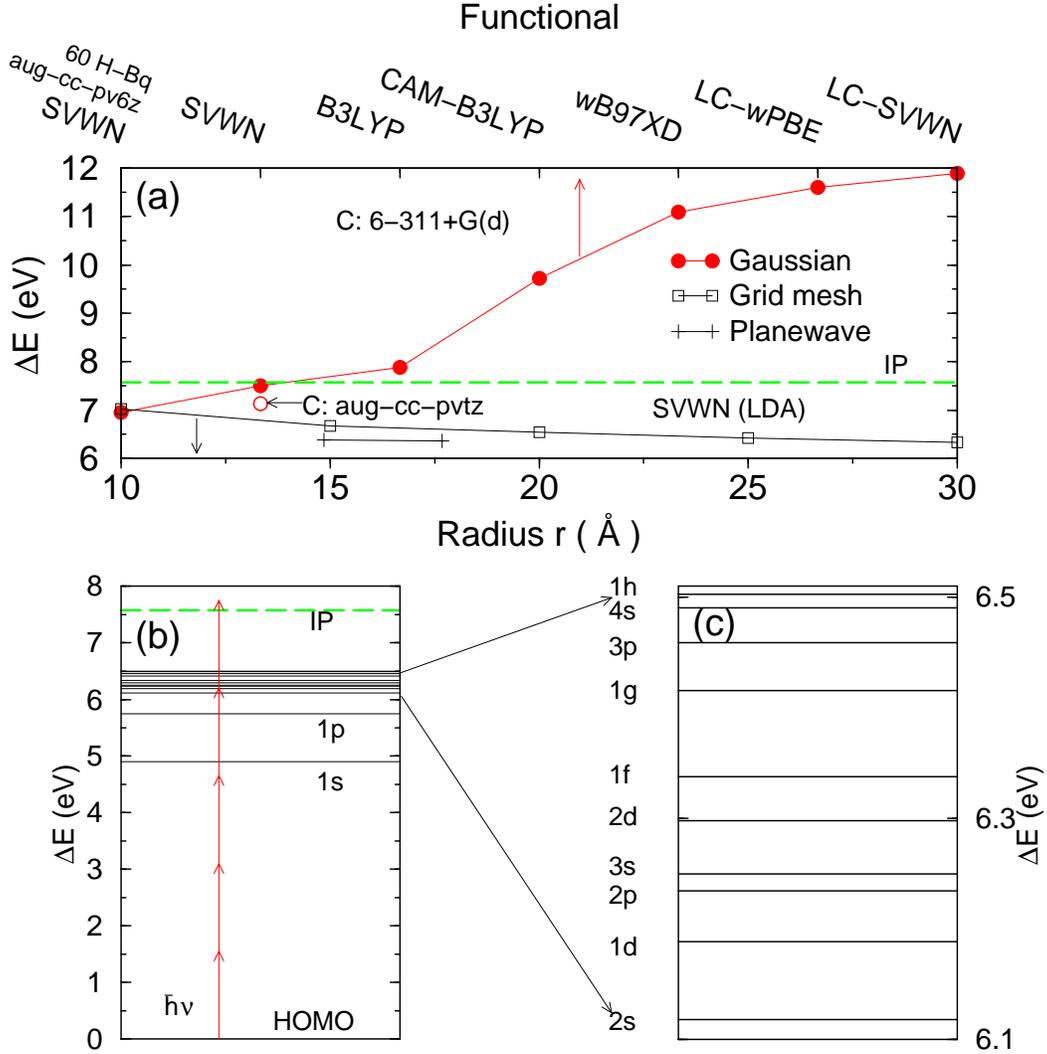}
\caption{ (a) Energy gap between the $1f$ SAMO and HOMO computed with
  different functionals and basis functions (top axis) and as a
  function of sphere radius (bottom axis). The filled circles
  represent the results computed with Gaussian09, while the empty
  squares are those obtained with Octopus (the pseudopotential and on
  the real grid mesh). We expect that in the infinite sphere limit,
  the gap settles down at 6.2 eV.  The VASP results (plus sign) are
  included for comparison, where the radius refers to the
  center-to-center distance of \ce.
(b) Energy spectrum of SAMOs referenced to the HOMO
  energy. The ionization potential energy is at 7.58 eV (see the thick
  dashed green line at the top). (c) Zoomed-in SAMO energy spectrum.  The
  $1f$ SAMO is between $2d$ and $1g$. The SAMOs form the doorway state
  for the ionization.  }
\label{fig2}
\label{energy}
\end{figure}

\begin{figure}
\includegraphics[angle=270,width=14cm]{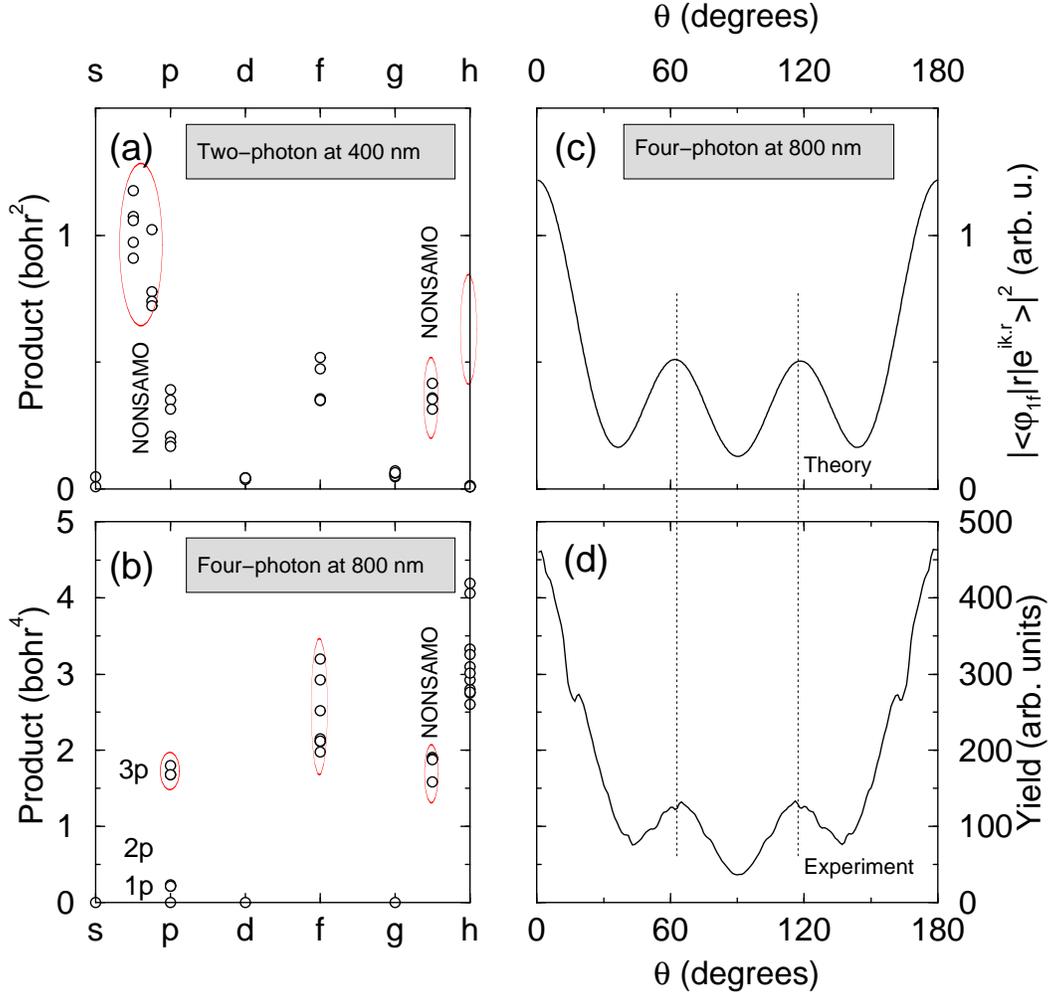}
\caption{ (a) Product of two transition matrix elements for the
  two-photon process at the 400-nm wavelength.  States that strongly
  contribute to the signal are highlighted with red ellipses.
  Although the nonSAMOs have a large transition matrix element, due to
  the off-resonance, their contribution is small. By contrast, 1s SAMO
  excitation is nearly resonant, so it dominates the photoelectron
  spectrum, consistent with the experimental findings.  (b) Products
  of four transition matrix elements for the four-photon process
  excited at 800 nm. In contrast to (a), four SAMOs ($2p,3p$, $1f$ and
  $1h$) and one nonSAMO contribute strongly. Since only a subset of
  SAMOs are excited, this casts an anisotropic image.  (c)
  Photoelectron angular distribution as a function of angle at 800
  nm. Two humps are separated by 56.5$^\circ$.  (d) Experimental
  results from Ref. \cite{li}. Excellent agreement with our theory is
  found.  }
\label{fig3}
\label{4photon}
\label{angular}
\end{figure}

To really image the $1f$ orbitals, we have to overcome two challenges.
First, since the $1f$ SAMO is seven-fold degenerate, would the final
image of $1f$ SAMO be completely washed out because of the spatial
average during the experiment, in particular as \cm spins rather
rapidly? This is an issue for the fluorescence spectrum in CO$_2$,
where the molecule has no preferred orientation. It has to be properly
aligned before it can be imaged \cite{xu}. Fortunately, we find that
the transition paths for seven $1f$ SAMOs are different, so their
products of transition matrix elements are not the same.  Thus, no spatial
average is necessary.  Numerically we find that they range from 3.2
to 2.1 $\rm \AA^4$ (see the circles at the $f$ SAMO in
Fig. \ref{angular}(b)). To understand the reason behind this, as an
example, we take two $1f$ orbitals with spherical harmonics $Y_{30}$
and $Y_{31}$, respectively. If the light is linearly polarized, for
$Y_{30}$, the transition path can be like $|\textrm
{HOMO}\ra\rightarrow Y_{00}\rightarrow Y_{10}\rightarrow
Y_{20}\rightarrow Y_{30}$, but for $Y_{31}$, $|\textrm {HOMO}\ra
\rightarrow |i\ra \rightarrow |j\ra \rightarrow Y_{11}\rightarrow
Y_{21}\rightarrow Y_{31}$, where $|i(j)\ra$ are two different
intermediate states. Therefore, the experimental geometry
(Fig. \ref{fsamo}) intrinsically differentiates among different $1f$
states, already found in atoms \cite{dixit}.  In \cm we find that the
dominant path or doorway state \cite{hertel92} is $|\textrm {HOMO}\ra
\rightarrow |1d\ra\rightarrow |2p\ra \rightarrow |2d\ra
\rightarrow|1f\ra $.

The second challenge is whether $1f$ SAMO can ever cast its
distinctive image on the final continuum state. We decide to compute
the angular distribution of the electron density for an $f$ orbital
with the spherical harmonic $Y_{30}$. The intensity is proportional to
the square of \be\la \psi(\vec{r})|\vec{p}|{\rm
  e}^{i\vec{k}\cdot\vec{r}} \ra = 4\pi i^3 \hbar \vec{k} Y^*_{30}
(\theta_k,\phi_k) \int_0^{\infty} dr R_{f}(r) j_3(kr) r^2
, \ee where $\theta_k$ and $\phi_k$ define the direction of the
wavevector $\vec{k}$ of the ejected electron, and a constant term due
to the orthogonality is not included. The angular distribution comes
from $Y^*_{30} (\theta_k,\phi_k)$. Employing the realistic
wavefunction of $1f$ SAMO in \ce, we compute the angular distribution
of the $1f$ SAMO.  The results are shown in Fig.
\ref{angular}(c). The maxima are at $0^\circ$ and $180^\circ$, along
the positive and negative directions on the $z$-axis, respectively
(see the $f$ SAMO in Fig. \ref{fsamo}).
 As the
angle rotates away from the $z$-axis, the intensity drops. We believe
that this produces the anisotropy seen in the experiment
\cite{li}.  The $1f$ SAMO has two distinctive
double humps, separated by 56.5$^\circ$. The unpublished experimental
photoelectron angular distribution by Li and coworkers \cite{li} is shown
in Fig. \ref{fig3}(d). The agreement between our theory and their
experiment is remarkable, where the shape, peak and hump locations are
identical.  We draw two dashed lines to highlight the locations of two
humps; and our estimated experimental angle separation between them is
54.4$^\circ$, in quantitative agreement with our theory within the
experimental error. Figure 4.6(f) of Ref.  \cite{li} looks more like
our predicted $1f$ SAMO (see Fig. \ref{fig1}).  This constitutes a
strong experimental support for our theory.

\section{Conclusion}
While much attention has been given to the diffuse spectrum in VMI,
recent experimental evidence shows a significant contribution from the
non-diffuse photoelectrons \cite{li}.  Here we demonstrate through a
series of carefully designed first-principles calculations that the
non-diffuse and anisotropic velocity image observed at 800 nm is
likely due to the excitation from the highest occupied molecular
orbital into the $1f$ SAMO.  We predict that the energy gap between
the highest occupied molecular orbital and $1f$ SAMO matches the
four-photon energy of 800 nm, and a peak at 0.2 eV will show up in the
photoelectron spectrum. In agreement with the prior study, two 400-nm
photons can excite 1\textit{s} SAMO from HOMO-2 nearly resonantly, although
nonSAMOs have a larger transition matrix product, but they are off
resonance by 0.6 eV. Since the 1s SAMO is spherical, the photoelectron
image is isotropic, in addition to the possible thermal and diffuse
electron emission.  Future experimental verification will have a
series of important theoretical consequences.  Which density
functional, LDA or GGA, is more accurate to describe the photoelectron
angular distribution and multiphoton emission? Which basis functions,
Gaussian diffuse functions or real grid mesh, are more accurate to
describe superorbitals? For the first time, we demonstrate that it is
the multiphoton excitation that greatly boosts the transition
probability into these superorbitals, a finding that is expected to
inspire new experimental and theoretical investigations.


\acknowledgments This work was supported by the U.S. Department of
Energy under Contract No.  DE-FG02-06ER46304 (GPZ).  We acknowledge
Hui Li, M. Kling, C. Lewis Cocke and Itzik Ben-Itzhak (Kansas State
University) for making the data from Ref. \cite{li} available to us.
We also thank E. Glendening (ISU) for the great help with the natural
bond orbital analysis in Gaussian09, S. W. Robey (NIST) for sending us
the Gaussian input file regarding Ref. \cite{steve}, H. Petek
(University of Pittsburgh) for their original paper \cite{feng},
technical support from the Gaussian team on the pseudoatom
applications, and D. Strubbe (MIT) for helpful communications with
Octopus.  HPZ and XSW were supported by the National Natural Science
Foundation of China (Contract Nos. 11274153, 10974081 and 10979017)
and National Key Projects for Basic Research of China (Contract
No. 2010CB923404). HPZ acknowledges support from China Scholarship
Council during her stay in the United States and thanks Indiana State
University for the hospitality under the exchange program.  Part of
the work was done on Indiana State University's Quantum clusters and
HPC clusters.  This research used resources of the National Energy
Research Scientific Computing Center, which is supported by the Office
of Science of the U.S.  Department of Energy under Contract
No. DE-AC02-05CH11231.  This work was performed, in part, at the
Center for Integrated Nanotechnologies, an Office of Science User
Facility operated for the U.S. Department of Energy (DOE) Office of
Science by Los Alamos National Laboratory (Contract DE-AC52-06NA25396)
and Sandia National Laboratories (Contract DE-AC04-94AL85000).

$^*$gpzhang@indstate.edu

\clearpage

\end{document}